# Classification Schemas for Artificial Intelligence Failures

## Peter J. Scott[1] and Roman V. Yampolskiy[2]

[1]Next Wave Institute, USA
[2]University of Louisville, Kentucky, USA
peter@humancusp.com, roman.yampolskiy@louisville.edu

## Abstract

In this paper we examine historical failures of artificial intelligence (AI) and propose a classification scheme for categorizing future failures. By doing so we hope that (a) the responses to future failures can be improved through applying a systematic classification that can be used to simplify the choice of response and (b) future failures can be reduced through augmenting development lifecycles with targeted risk assessments.

Keywords: artificial intelligence, failure, AI safety, classification

## 1. Introduction

Artificial intelligence (AI) is estimated to have a \$4-6 trillion market value [1] and employ 22,000 PhD researchers [2]. It is estimated to create 133 million new roles by 2022 but to displace 75 million jobs in the same period [6]. Projections for the eventual impact of AI on humanity range from utopia (Kurzweil, 2005) (p.487) to extinction (Bostrom, 2005). In many respects AI development outpaces the efforts of prognosticators to predict its progress and is inherently unpredictable (Yampolskiy, 2019). Yet all AI development is (so far) undertaken by humans, and the field of software development is noteworthy for unreliability of delivering on promises: over two-thirds of companies are more likely than not to fail in their IT projects [4]. As much effort as has been put into the discipline of software safety, it still has far to go.

Against this background of rampant failures we must evaluate the future of a technology that could evolve to human-like capabilities, usually known as *artificial general intelligence* (AGI). The spectacular advances in computing made possible by the exponential hardware improvements due to Moore's Law (Mack, 2011) balanced against the unknown required breakthroughs in machine cognition make predictions of AGI notoriously contentious. Estimates of how long we have before AGI will be developed range over such widely varying timelines (Sotala & Yampolskiy, 2015) that researchers have taken to metaanalysis of the predictions through correlation against metrics such as coding experience of the predictors [5].

Less contentious is the assertion that the development of AGI will inevitably lead to the development of ASI: *artificial superintelligence*, an AI many times more intelligent than the smartest human, if only by virtue of being able to think many times faster than a human (Vinge, 1993). Analysis of the approach of confining a superintelligence has concluded this would be difficult (Yampolskiy, 2012) if not impossible (Yudkowsky, 2002). Many of the problems presented by a superintelligence resemble exercises in international diplomacy more than computer software challenges; for instance, the *value alignment problem* (Bostrom, 2005) (described therein as the "value loading problem") of aligning AI values with humans'.

## 2. Definitions

*Artificial intelligence* is a shifting term whose definition is frequently debated. Its scope changes depending upon the era: during an "AI Winter" (Crevier, 1993) many fewer vendors are willing to identify their products as AI than during the current period of myriad AI technologies clogging the "peak of inflated expectations" in the Gartner Hype Cycle. [6]

*Failure* is defined as "the nonperformance or inability of the system or component to perform its expected function for a specified time under specified environmental conditions." (Leveson, 1995). This definition of failure as an event distinguishes it from an *error,* which is a static condition (or state) that may lead to a failure.

*Cybersecurity* has been defined as "the organization and collection of resources, processes, and structures used to protect cyberspace and cyberspace-enabled systems from occurrences that misalign *de jure* from *de facto* property rights." (Craigen, Diakun-Thibault, & Purse, 2014). AI Safety has been defined as an extreme subset of cybersecurity: "The goal of cybersecurity is to reduce the number of successful attacks on the system; the goal of AI Safety is to make sure zero attacks succeed in bypassing the safety mechanisms." (Yampolskiy, 2016).

*Intelligence* definitions converge toward the idea that it "… measures an agent's ability to achieve goals in a wide range of environments." (Legg & Hutter, 2007). We do not present this definition with any intention of defining AI by applying the "artificial" modifier to this one. Rather, this definition will be used to judge whether a software failure is instructive in the extent to which it was applying (accidentally or intentionally) intelligence in even the narrowest sense, since such application could extend to a more powerful AI.

## 3. AI Failure Classification

It is precisely because of the volatile definition of AI that we must cast a wide net in what we use for examples of AI failures, because what is classified as AI today will likely be given a less glamorous title (like "machine vision") once it becomes commonplace. As AI pioneer John McCarthy put it, "As soon as it works, no one calls it AI any more." [7] Where some of our examples, therefore, may appear to be indistinguishable from failures of software that has no particular claim to the label of artificial intelligence, they are included because they are close enough to AI on the software spectrum as to be indicative of potential failure modes of AI.

### 3.1 Historical Classifications

Neumann (Neumann, Computer-Related Risks, 1994) described a classification for computer risk factors (see table 1).

| Problem sources and examples |
|---|
| Requirements definition, omissions, mistakes |
| System design, flaws |
| Hardware implementation, wiring, chip flaws |
| Software implementation, program bugs, compiler bugs |
| System use and operation, inadvertent mistakes |
| Wilful system misuse |
| Hardware, communication, or other equipment malfunction |
| Environmental problems, natural causes, acts of God |
| Analysis of concepts, design, implementation, etc |
| Evolution, maintenance, faulty upgrades, decommission |

Table 1: Neumann Computer Risks sources and examples

We find this list too broad in some respects and too narrow in others to be useful for our purposes. Hardware factors are outside the scope of this paper (and are increasingly irrelevant as software becomes more platform-independent and mobile); software factors need greater elaboration. Neumann and Parker (1989) listed classes of computer misuse techniques (see figure 1).

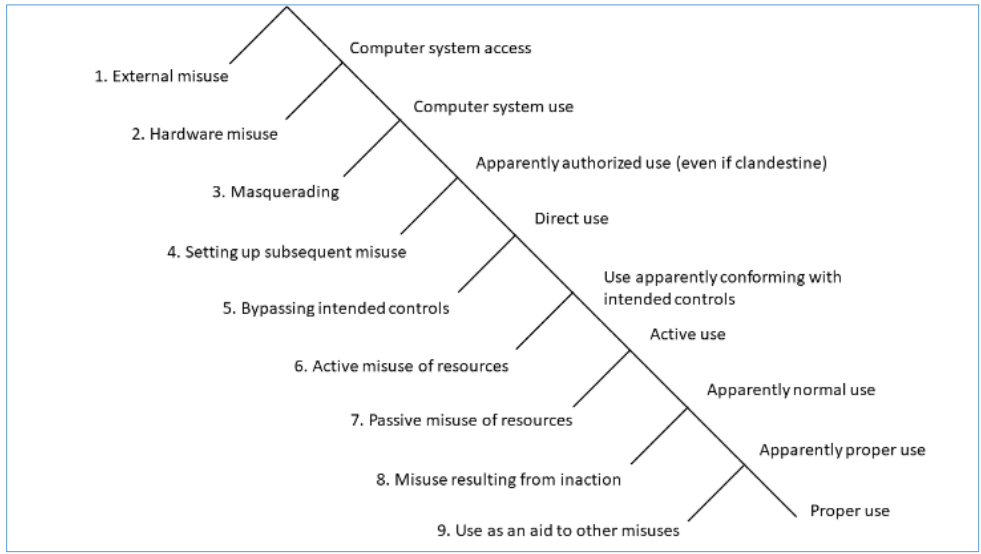

Figure 1: Neumann and Parker Computer misuse technique classes

Despite the tree structure, this represents a system of descriptors rather than a taxonomy in that a given misuse may involve multiple techniques within several classes. The leftward branches all involve misuse; the rightward branches represent potentially acceptable use–until a leftward branch is taken. However, the term "misuse" implies deliberate agency and thereby ignores a multitude of failure modes that stem from accidental oversights.

*3.2 AI Failure Classification Dimensions*

Here we modify and extend earlier work by Yampolskiy (2016) in classifying AI risk factors. Hollnagel (2014) deconstructs safety in the steps of *phenomenology* (observables), *etiology* (causes), and *ontology* (nature). We address each of these steps in proposing the following dimensions as useful classification criteria for AI failures:

- Consequences (phenomenology)
- Agency (etiology)
- Preventability (ontology)
- Stage of introduction in the product lifecycle (phenomenology and ontology)

Each will be denoted with a 2- or 3-letter code that we will tag our examples with.

3.2.1 *Consequences* may be considered on the scale of human aggregation on which they can occur. (See table 2.)

| Human Aggregation Scale | Consequences | | | | | |
|---|---|---|---|---|---|---|
| | **Physical** | **Mental** | **Emotional** | **Financial** | **Social** | **Cultural** |
| Individual | CIP | CIM | CIE | CIF | | |
| Corporation | | | | CCF | | CCC |
| Community | | | | CYF | CYS | CYC |

Table 2: AI Failure Consequences at Human Aggregation Levels

*Individuals* can range in number from one to every member of the human race; the grouping will be used to denote at what type of aggregation the action of the failure was aimed rather than the number of instances affected. *Corporations* are legal structures for doing business, of any size. *Communities* are groupings of people organized for purposes other than business and range from families to nations.

- *Physical* consequences occur to individuals and may range from inconvenience to loss of life.
- *Mental* consequences occur to individuals and include the alteration of mental states such as beliefs, with concomitant changes in behavior. For instance, the purpose or effect of "fake news" is to cause such changes. (Lazer, et al., 2018).
- *Emotional* consequences occur to individuals and include depressive states resulting from AI incidents with physical or mental consequences, and AI usurping roles that people have assumed to be unassailable.
- *Financial* consequences occur to individuals, corporations, and communities.
- *Social* consequences are the modifications of behavior of systems or organizations of people.
- *Cultural* consequences are the modifications of an organization or grouping's vision, values, norms, systems, symbols, language, assumptions, beliefs, and habits (Needle, 2004).

Consequences are not necessarily negative, or may be negative in some respects while being positive in others. A superintelligence that enslaved humans in boot camps might keep them in optimal physical condition but pessimal emotional state.

3.2.2 The *agency* of a failure is the degree of human intentionality in its origin or propagation (see table 3).

| Agency | Code |
|---|---|
| Accidental | AA |
| Negligent | AN |
| Intentional | AI |
| Malicious | AM |

Table 3: AI Failure Levels of Agency

An *accidental* failure is one that was not foreseen and could not reasonably have been foreseen. We are departing slightly from the customary engineering definition of 'accident' here in order to draw a more useful distinction. Leveson (1995) defines 'accident' as "An undesired and unplanned (but not necessarily unexpected) event that results in (at least) a specified level of loss." Thus automobile accidents are foreseeable but neither expected nor desired. We prefer instead to define a *negligent* failure as one that was not foreseen but could (and perhaps should) have been foreseen.

An *intentional* failure is one deliberately caused, but not with malicious intent, possibly with the intent of causing a more benign effect than what actually resulted. A *malicious* failure is one that was initiated with the intention of causing deleterious effects, whether they were specifically the effects that actually resulted or others. No connection with legal definitions of these terms should be inferred from their attribution to specific events.

3.2.3 Levels of agency are independent of the *degree of preventability* (see table 4).

| Degree of Preventability | Code |
|---|---|
| Trivially preventable | PT |
| Preventable with some difficulty | PS |
| Preventable with great difficulty | PD |
| Unpreventable | PU |

Table 4: AI Failure Degree of Preventability

Some failure modes of superintelligences are forecast by some authorities to be unpreventable: "[W]e have seen enough to conclude that scenarios in which some machine intelligence gets a decisive strategic advantage are to be viewed with grave concern." (Bostrom, 2005) (p.154).

3.2.4 A common taxonomy for computer system errors is the software development *lifecycle stage* (see table 5); it is often asserted that the cost of fixing an error at each stage is ten times the cost of fixing it in the previous stage (Dawson, Burrell, Rahim, & Brewster, 2010).

| Lifecycle Stage | Code |
|---|---|
| Concept | LC |
| Design | LD |
| Development | LE |
| Testing | LT |
| Operation | LO |
| Decommissioning | LG |

Table 5: Software Development Lifecycle Stages

We add in the less commonly included stages of concept (was it a good idea to do this in the first place?) at the beginning, and decommissioning (what are the problems caused by getting rid of the product) at the end. A superintelligence might be highly resistant to decommissioning. [10]

## 4. AI Failures

With these dimensions in mind we now examine various reported and hypothesized failures. Note that there is an unavoidable degree of subjective variability in the classifications of preventability and agency.

### *4.1 Reported Failures*

Whereas Yampolskiy (2018) enumerated several dozen failures in a timeline that highlighted an exponentially increasing frequency and severity, nearly all of the examples we cite here occurred within the 2016-2019 period and so a chronological ordering would not be illuminating. We will therefore place them instead within a more narrative structure.

The most recognizable and straightforward class of failures result in physical injury to humans, going back to the classic Therac-25 radiation therapy overdose cases (Rawlinson, 1987) (CIP, AN, PS, LD, LE, LO). When an Amazon warehouse robot accidentally punctured a container of bear spray [9] (CIP, AN, PS, LT) it was a more benign outcome of an industrial accident than when a Chinese factory worker was impaled with ten foot-long spikes [10] (CIP, AN, PT, LD). But these and other more fatal accidents with industrial robots going back at least to 1984 when an operator was killed by a 2,500 lb robot that came behind him with no warning (Fuller, 1984) (CIP, AN, PS, LD) indicate lack of consideration for humans sharing the same location as machines. A car production plant robot grabbed a worker instead of a part and crushed him against a metal plate, killing him [] (CIP, AN, PS, LD).

Incidents of cars in semi-autonomous operation causing fatalities include an Uber incorrectly classifying a pedestrian as a false positive match because too many reactions to actual false positives resulted in a jerky ride [12] (CIP, AN, PS, LT), and a Tesla crashing after requesting driver intervention [13] (CIP, CCF, AA, PD, LE).

In medicine, IBM's Watson recommended 'unsafe' cancer treatments [14] (CIP, CCF, AA, PD, LT, LO), and a study of 14 years of robotic surgery concluded that "a non-negligible number of [preventable] technical difficulties and complications are still being experienced during procedures." (Alemzadeh, Raman, Leveson, Kalbarczyk, & Iyer, 2016) (CIP, CCF, AA, PS, LD).

AI accidents may result in direct financial loss. The May 2010 "Flash Crash" resulted in the Dow Jones Industrial Average dropping about 9% for 36 minutes and resulted from program trading algorithms being inadequately prepared to deal with large volumes of strategically-placed trades which themselves were computer-mediated malice [16] (CIF, CCF, AA, AM, PD, LD, LT). Remediation efforts did not prevent more flash crashes in 2015 [17].

A major concern in the application of AI is privacy. Consumer devices connected to corporate clouds of identity data come under scrutiny, especially when, for instance, an Amazon Alexa node recorded a private conversation and sent it to a random contact [15] (CIE, AA, PS, LT, LO), or an iPhone bug allowed users to listen on others' conversations via FaceTime [18] (CIE, AA, PS, LT). In some cases, the technology facilitated a casual violation of privacy such as when Uber users' locations and identities were displayed on a screen at a launch party [19] (CIE, AI, PT, LC).

Privacy violations carry more serious consequences when they become misidentifications. The ACLU demonstrated that when they showed that Amazon facial recognition would flag certain members of Congress as wanted criminals [20] (CYF, CYS, AN, PS, LD). A lack of training data (and implicit bias) resulted in facial recognition systems being unable to see black people [21] or tagging them as gorillas [22] (CIE, CYS, AN, PD, LD). Facial recognition used by police in the United Kingdom has been recorded making many false positive identifications [23] (CIE, CIF, CYF, CYS, AN, PS, LT, LO). And in China, facial recognition systems deployed for automated misdemeanor ticketing publicly shamed a woman as a jaywalker when mistaking her photo on the side of a bus for the woman herself [24] (CIE, CIF, AN, PD, LE) and a driver was ticketed for using a cellphone when he was actually scratching his face [25] (CIE, CIF, AN, PD, LE). Traffic cameras in New Orleans ticketed parked cars for speeding [26] (CIF, AN, PS, LD, LO). A man was falsely arrested after systems at Apple misidentified him as stealing from its stores [27] (CIP, CIM, CIE, CIF, CCF, AA, PS, LE).

Not all misidentifications result in such obvious harm. Artist Tom White specializes in creating abstract (and very unarousing) art that is flagged as unacceptable nudity by social media AI. [28] This machine myopia indicates that the development of useful image censorship is not yet realized and some inoffensive art is suppressed. (CIM, CIF, AA, PD, LD).

Implicit misidentification by category is *bias,* another topic of great concern in AI development. With good reason: a report concluded that AIs trained on hiring decisions would replicate or amplify human bias [29], Amazon's hiring AI turned out to be sexist [30], and the COMPAS system used in Wisconsin to predict recidivism was biased against blacks [31] (CIE, CIF, CYC, AA, PD, LE). Just as human bias often results from inadequate exposure to diversity, AI bias often arises from the same cause. An attempt to use AI to objectively judge an online international beauty contest without human bias failed when only one of 44 winners it chose had dark skin, prompting speculation that this was due to the training database having few dark faces [32]. And the New Zealand automated passport application checking system rejected an Asian applicant's photograph, claiming that "Subject's eyes are closed" [33]. A study demonstrated that implicit race and gender biases in training corpora flowed through into AIs trained on those corpora (Caliskan, 2017)

In the hands of an authoritarian regime, AI can create environments prompting comparisons with Orwell's *1984.* Nowhere is this more apparent than in China, which has embraced facial recognition on a large scale [34]. AI there blocks mention of the Tiananmen Square massacre on social media [35] (CYS, AI, PT, LC). While this software is being used to create exactly its intended effect, we label this a failure because it has consequences many western observers would consider to be socially harmful. China has a "social credit" scoring system reminiscent of a Black Mirror episode (Wright, 2016), linked to social media and consumer systems such as Sesame Credit [36], that will ban people from certain venues like flights and hotels for poor scores, which may be incurred by undesirable behaviour such as buying video games (CYC, AI, PT, LC). Some commentators speculate that this will have consequences in health care [37]. Also in China, AI is being used to grade school papers [38], with some good writing being given poor marks (CYS, AI, PD, LO). And AI is used to monitor the moods of workers [39] and the attention paid by children in class [40], with the most attentive being rewarded (CIM, CYC, AI, PT, LC).

In the West the dangers are more nascent. Researchers at the University of Pennsylvania demonstrated that textual analysis of an individual's Facebook posts could predict 21 different medical conditions such as diabetes (Merchant, Asch, Crutchley, Ungar, & Guntuku, 2019). Others showed that AI was better than people at determining sexual orientation from a photograph (Wang & Kosinski, In Press), while a third group determined that AI could detect certain genetic diseases from faces [41]. A Department of Homeland Security program predicts which flyers are potential terrorists [42] from demographic and travel data alone, and if those travellers make it to the European Union they may face an AI-powered lie detection system at the border [43]. The startup Faception claims its software can predict personality traits such as pedophile or poker player from facial image analysis, causing one commentator to liken it to phrenology [44]. A person's gait can be used to identify them [45]. Two systems supply "predictive policing" systems that, inviting a comparison with the movie *Minority Report,* suggest where crime is likely to occur [46] [47]. Companies exploit human psychology to get our attention [48], the US military studies how to influence Twitter users [49], and the Pentagon wants to predict protests against the US President via social media surveillance [50].

As Yampolskiy (2016) pointed out, "An AI designed to do X will eventually fail to do X," codified as the *Fundamental Theorem of Security:* There is no such thing as a 100% secure system. In all the examples in the previous paragraph the latent failures are the ones implied by this theorem, with their concomitant risks.

The consequences of *misinformation* spread by AI include of course "fake news," such as that attributed to Cambridge Analytica [52], assiduously spread by social media [53]. and "deep fake" videos [54], which could be used to automate blackmail at scale [55] (CIM, CYS, AM, PU, LO)

A class of incidents illustrates that much AI is not yet mature. A hardware design bug allowed memory protection violations in years' worth of Intel chips [56] (CCF, AA, PD, LD). And Microsoft's Tay chatbot became racist within hours of being deployed to learn from other Twitter users [57] (CYS, AA, PS, LC/LD). Some aspects of this immaturity are fundamentally brittle; for instance, when a digital exchange lost $137 million because the one person holding the master password died [58], or when bots tasked with maintaining Wikipedia fought with each other for years [59]. Deep reinforcement learning fails more often than admitted [60].

Intentional misuse spans many incidents; to cite two, smart scooters for hire were hacked to display obscene messages and be used without payment [61] (CCF, AM, PS, LE) and Domino's Pizza affiliation app was fooled into granting points by fake pictures of pizza [62] (CCF, AM, PD, LC).

Some "backfire" events result in damage to the AI industry through overreaching or misrepresentation. For instance, a preternaturally capable healthcare AI called "Zach" in New Zealand was suspected to be a person in disguise [63]. And the Sophia robot attracts a degree of adulation far beyond its real capabilities [64]. The threat here is to the reputation of AI and its community (CYF, CYC, AI, PU, LO).

AI that is unintentionally insensitive also damages its own reputation, such as the AI that thought that house burning down was "spectacular" [65] and Paypal's virtual assistant which insensitively replied, "Great!" when someone told it, "I got scammed" [66] (CIE, AA, PS, LE). The Starbucks shift-scheduling software was also insensitive when it optimized for hour-by-hour business needs but assigned workers to unpredictable and erratic schedules [67] (CIP, CIF, CCC, AN, PT, LD). AI that is trusted without verification may not live up to that trust, such as when a model used to grade the "value-add" imparted by New York City teachers was found to generate essentially random results [68] (CIM, CIE, CIF, AN, PS, LT). A corporate employment workflow system was unstoppable in terminating an employee erroneously flagged as superfluous [69]; after three weeks spent fixing the error he declined to return to the firm (CIE, CIF, CCF, AA, PS, LD).

Some failures are so benign on the surface that many casual observers would classify them as cute behaviour rather than failures. When a robot (with smiley face to boot) on the International Space Station stopped obeying astronauts [70] the parallels with HAL 9000 of *2001: A Space Odyssey* were so irresistible as to obscure the real risks of a computer failure in a critical environment. Apple's Siri's initial response to the request "Call me an ambulance" was to refer to the user thereafter as "ambulance" [71] (CIP, AA, PS, LE). When a text generator created weird descriptions of Bitcoin [72], and an AI's predicted YouTube pornography searches [73], the results were so funny as to be equally disarming (CYC, AA, PS, LD). A trivial typo in the code for a game agent made it much easier to beat than it should have been [74] (CIM, AA, PT, LT). A Roomba spread dog poop all over a house [75] (CIP, CIF, AA, PS, LE). A sign printed in Welsh translated to "I am not in the office at the moment. Send any work to be translated." [76] (CYC, AN, PT, LO). The "swarm intelligence" UNU failed to predict the results of the Kentucky Derby the second time around after previously winning the superfecta. [77] (CIF, AA, PD, LE). A neural network hallucinated sheep in images where there were none, or mislabelled them when they were placed in (admittedly unusual) locations [78] (CIM, AA, PS, LE). And in a story guaranteed to get more laughs than fears of AI failure, Alexa devices were alleged to be spontaneously laughing [79].

Behavior that is also perceived as "cute" in the sense of "look at how smart my child is," can be more concerning because it indicates just how creative AI can be in solving problems with solutions that eluded humans. AIs "cheat" at games by finding loopholes in the rules or unintended back doors in the implementation [80]. One AI invented (or rediscovered) steganography in order to meet its goals [81]. And GPT2, a text generator developed by OpenAI, an organization dedicated to open sourcing AI to ensure its safety, was deemed to be so good at what it did that it would be too dangerous to publish its code [82].

*4.1.3 Genetic Algorithms*

Genetic algorithms can be so innovative at "breaking the rules" [83] and (Lehman, Clune, Misevic, & Adami, 2018) that they check every category of failure classification, suggesting a path towards unbounded risk.

- They can exploit misfeatures or bugs in their environment, such as when in the developmental stages of the NERO video game, players' robots evolved a wiggling motion that allowed them to walk up walls rather than solve the obstacles "properly" by walking around the walls (Stanley, 2005), or when in a capstone project for a graduate level class, students were required to make a a five-in-a-row Tic-Tac-Toe game played on an infinitely large board. One submission's algorithm evolved to request non-existent moves that were extremely far away, leading to an automatic win since the other players system would crash (Moriarty, 1997).

- They can "cheat" by exploiting loopholes in the rules of their goals, such as in an experiment that involved organisms navigating paths, when one organism created an odometer to allow it to navigate the path precisely and earn a perfect score (Grabowski, 2013), or when an attempt to create creatures that could evolve swimming strategies resulted in them learning that by twitching their body parts rapidly, they could obtain more energy that let them swim at unrealistic speeds (Sims, Evolving virtual creatures, 1994).

- They can reinvent, to their creators' surprise, capabilities of biological organisms, such as in an experiment where robotic organisms had to find foods or poisons that were both represented by red lights and could use blue lights to communicate with other robots, the organisms evolved in surprising ways that resembled mimicry and dishonesty in nature (Mitri, 2009), or when a digital evolution model that was initially thought to have been a complete failure, was discovered to have reproduced the biological concept Drake's rule without having been told to do so (Hindré, 2012).

- They can improvise novel solutions to their assigned tasks, such as when 3-D creatures that could run, walk, and swim were gauged by a fitness function of average ground velocity, which resulted in creatures that were tall and rigid, falling over and using their potential energy to gain high velocity (Sims, 1994b), or when a robot arm was programmed to interact with a small box on a table, but the gripper was broken, resulting in the robot hitting the box with the gripper in a way that would force the gripper to hold the box firmly (Ecarlat, 2015).

- They can creatively exceed their goals, such as the artificial life system Tierra, not expected to evolve higher life forms for years, created complex ecological systems on the first successful run (Ray, 1992), or when robots that were designed to detect and travel to a light source evolved a spinning behavior that was more efficient than the expected Braitenberg-style movement (Watson, 2002).

*4.2 Hypothetical Failures*

A video producer depicted a fictional future where an artificial superintelligence charged with copyright enforcement hacked people's brains with nanotechnology to correct violations [84] (CIM, CYC, AI, PD, LO). It was demonstrated that a DNA sequencer could be hacked through (currently non-existent) flaws in a compression algorithm [85] (CIP AM, PD, LE).

Most shows that explore AI failure develop a theme epitomized by *Terminator* series: a massive AI becomes self-aware and attempts to destroy humanity. (CIP, CIE, CIF, CCF, CYF, CYS, CYC, AN, AI, PD, LC, LO). Variations include *Colossus: The Forbin Project,* where the AI imprisons humanity to end conflict (CIM, CIE, CYS, CYC, AN, AI, PD, LC, LO), the same goal as the AI VIKI in the movie *I, Robot* and the robots in Jack Williamson's novelette "With Folded Hands" (Williamson, 1947). One of the least apocalyptic failures was explored in the 2013 film *Her,* where virtual assistant AIs have unforeseen intimate relationships with many humans who are largely changed for the better (CIE, AA, PD, LD).

### 4.2.1 Broad Classifications of Future Failure Scenarios

Another classification for failures can be applied to future scenarios.

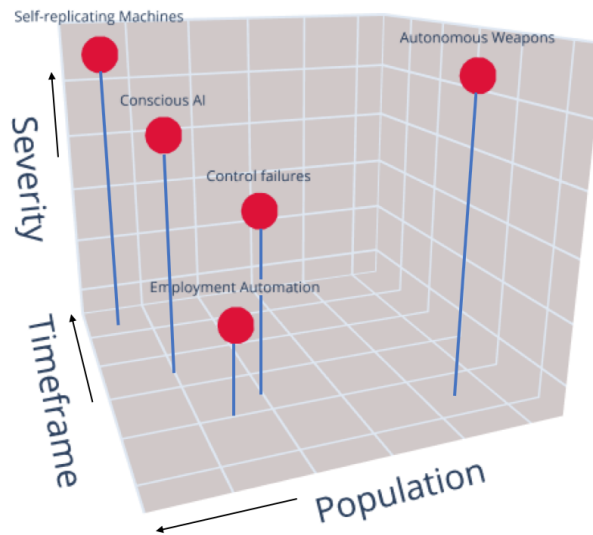

Figure 2: Near- and Far-Term AI Failure Scenarios

Figure 2 depicts the severity and scale (number of individuals affected) of broadly classified failure scenarios. In chronological order these are:

1. Autonomous weapons, which currently mostly fall into the 'lethal' category [86], [87].
2. Employment automation: The potential segment of the population made jobless through AI automation.
3. Control failures: AI of sufficient complexity and power that bugs cause catastrophes.
4. Conscious AIs: Control failures in AGIs or AIs that are so complex that their behavior is most usefully categorized as 'conscious.'
5. Self-replicating machines: Embodied AIs that can create copies of themselves from raw materials in the environment.

The scenario of 'Conscious AIs' merits some elaboration. Whether an AI is actually conscious is going to become an increasingly difficult and contentious question to answer, but this scenario does not depend on the answer. The "apparently conscious" AIs in this category are ones that, whether they are conscious or not, will be doing such a good impression of consciousness that it would be more productive to think of them that way than to apply traditional computer science methods to them. We will have reached this stage when the field of "AI psychiatry" comes into existence.

The chart is not to scale; these are qualitative assessments intended to provoke and inform strategic planning. While some of these labels are apocalyptic, we are motivated by considering Normal Accident Theory (Perrow, 2011) and Maas' application to AI: "At their extreme, unexpected interactions between competing systems, especially in cyberspace, could cause unexpected escalation—a 'flash war', analogous to the algorithmic flash crashes observed in the financial sector." (Maas, 2018)

## 5. Responses

There are various responses to these failures and risks. Several address privacy. "Differential privacy" masks individual data in Big Data collections. [86]. The Myelin framework preserves privacy in trusted hardware enclaves [89]. Another approach encrypts data before using it to train neural networks without loss of capability [90]. The Data Selfie browser add-on shows leakage of personal data [91]. Another program confuses ad tracking by clicking on every ad in the background [92]. A Facebook container isolates your Facebook activity from everything else you do [94] and a program creates search noise to drown out your actual searches [93].

Defenses are being developed against hacking image recognition networks through microchanges [95].

## 6. Conclusions

While we have not made recommendations as to how to address AI failures in each category of the dimensions we have presented, we hope that this classification scheme will make the development of remediation approaches easier.

The importance of this effort may be extrapolated from Leveson's observation that "The design of the automated system may make the system harder to manage during a crisis." (Leveson, 1995). Noting that this was true of the state of the art in 1995, we are concerned with how systems that are not just far more automated but autonomous may also be far harder to manage during a crisis. The more complex a system becomes, the larger the task X that may be assigned to that system, and so the larger the consequences of the system failing to do X. Today, a humor-generating system writes a joke that isn't funny; tomorrow, employee screening software will hire the wrong people, next week, a system designed to protect a national power grid from cyberattack will fail to do that, etc. Observe that AI systems that perform common human-centric tasks such as image recognition do so in ways that are unrelated to how humans perform those tasks, and are consequentially easily fooled by near-invisible changes [96]; that furthermore AI can operate on completely alien concepts such as the "opposite" of an image to show, e.g., the opposite of a cat [97]. These examples indicate that AI systems used to perform complex human-like tasks will have extremely unpredictable failure modes.

Some people in the AI community view these discussions as scaremongering that impedes the development of AI; to them we quote William Bogard chronicling the Bhopal chemical plant tragedy:

"We are not safe from the risks posed by hazardous technologies, and
any choice of technology carries with it possible worst-case scenarios that we must take into account in any implementation decision. The public has the right to know precisely what these worst-case scenarios are and participate in all decisions that directly or indirectly affect their future health and well-being. In many cases, we must accept the fact that the result of employing such criteria may be a decision to forego the implementation of a hazardous technology altogether." (Bogard, 1989)

**Notes**

1.

   https://www.mckinsey.com/~/media/mckinsey/featured%20insights/artificial%20intelligence/notes%20from%20the%20ai%20frontier%20applications%20and%20value%20of%20deep%20learning/notes-from-the-ai-frontier-insights-from-hundreds-of-use-cases-discussion-paper.ashx
2. https://www.bloomberg.com/news/articles/2018-02-07/just-how-shallow-is-the-artificial-intelligence-talent-pool
3. http://www3.weforum.org/docs/WEF_Future_of_Jobs_2018.pdf
4. https://www.batimes.com/articles/the-impact-of-business-requirements-on-the-success-of-technology-projects.html
5. https://reducing-suffering.org/predictions-agi-takeoff-speed-vs-years-worked-commercial-software/
6. https://www.ciodive.com/news/gartner-serves-up-2018-hype-cycle-with-a-heavy-side-of-ai/530385/
7. https://cacm.acm.org/blogs/blog-cacm/138907-john-mccarthy/fulltext
8. https://www.youtube.com/watch?v=Wy4EfdnMZ5g
9. https://www.engadget.com/2018/12/06/amazon-workers-hospitalized-robot
10. https://www.thesun.co.uk/news/7954270/factory-robot-malfunctions-and-impales-worker-with-10-foot-long-steel-spikes/
11. https://www.theguardian.com/world/2015/jul/02/robot-kills-worker-at-volkswagen-plant-in-germany
12. https://arstechnica.com/tech-policy/2018/05/report-software-bug-led-to-death-in-ubers-self-driving-crash/
13. https://www.washingtonpost.com/news/dr-gridlock/wp/2018/04/01/ntsb-unhappy-with-tesla-release-of-investigative-information-in-fatal-crash
14. https://www.statnews.com/2018/07/25/ibm-watson-recommended-unsafe-incorrect-treatments/
15. https://www.fox13memphis.com/news/trending-now/amazon-alexa-recorded-private-conversation-sent-it-to-random-contact-woman-says/755720160